\documentclass{sig-alternate-05-2015}

\usepackage{newtxtext}
\usepackage{amssymb}
\usepackage{amsmath}
\usepackage{graphicx}
\usepackage[hide]{notes-alt}
\usepackage[ruled,vlined,linesnumberedhidden]{algorithm2e}
\usepackage{booktabs}
\usepackage{url}
\usepackage{listings}
\usepackage{color}
\usepackage[usenames,dvipsnames,svgnames,table]{xcolor}

\newdef{definition}{Definition}

\CopyrightYear{2016} \setcopyright{acmcopyright}
\conferenceinfo{CHIIR'16,}{March 13--17, 2016, Carrboro, North Carolina, USA.}
\isbn{978-1-4503-3751-9/16/03}\acmPrice{\$15.00}
\doi{http://dx.doi.org/10.1145/2854946.2854959} 

\begin{document}
%




\title{History by Diversity: Helping Historians search News Archives}

 \numberofauthors{1}
 \author{
 \alignauthor
 Jaspreet Singh, Wolfgang Nejdl, Avishek Anand\\
        \affaddr{L3S Research Center, Leibniz Universit\"at Hannover.}\\
        \affaddr{Appelstr. 9a}\\
        \affaddr{30167 Hanover, Germany}\\
        \email{\{singh,nejdl,anand\}@L3S.de}
 }

\maketitle

\maketitle \begin{abstract} 

Longitudinal corpora like newspaper archives are of immense value to
historical research, and time as an important factor for historians
strongly influences their search behaviour in these archives. While
searching for articles published over time, a key preference is to
retrieve documents which cover the important aspects from important
points in time which is different from standard search behavior. To
support this search strategy, we introduce the notion of
a \emph{Historical Query Intent} to explicitly model a historian's
search task and define an aspect-time diversification problem over
news archives. 

We present a novel algorithm, \textsc{HistDiv}, that
explicitly models the aspects and important time windows based on a
historian's information seeking behavior. By incorporating 
temporal priors based on publication times and temporal expressions, 
we diversify both on the aspect and temporal dimensions. We test our methods by
constructing a test collection based on \emph{The New York Times
  Collection} with a workload of 30 queries of historical intent
assessed manually. We find that \textsc{HistDiv} outperforms all
competitors in subtopic recall with a slight loss in precision. 
We also present results of a qualitative user study to determine 
wether this drop in precision is detrimental to user experience. 
Our results show that users still preferred \textsc{HistDiv}'s ranking.


\end{abstract}




\section{Introduction}
\label{sec:intro}

Newspaper articles encode history as it happens by capturing events and their immediate impact on society, politics, business and other important spheres. These are of immense value to historians, sociologists, and journalists who rely on fairly reliable, accurate and time-aligned information sources. Specifically for historians, whose desired corpus of study is an archive, browsing and searching such archives has emerged as an important aspect in their research~\cite{tibbo2003primarily}. Consequently, designing access methods and retrieval models tailored to their search patterns and information need is an important problem.

The information seeking behavior of a historian is slightly different from the traditional user search behavior, for which classical retrieval tasks are designed, in two respects. First, historians are interested in obtaining an overview of the topic they wish to research in order to contextualize results. They desire to look at relevant results from \emph{important subtopics} from the most \emph{relevant time points of interest}. Currently, this is realized by issuing an underspecified broad query on the topic and then trying to identify relevant articles from important subtopics by applying various filters which are time, source, region or domain-based. Secondly, the major preoccupation of historians is in finding \emph{primary sources} of information (accounts/reports/documents made by an observer of an event). Secondary information sources (accounts made in retrospect) are also important, however they are intrinsically used to identify primary information sources.

Consider a historian interested in Rudolph Giulaini, a U.S. Republican politician, in the period between 1987 and 2007. Giuliani started out as an attorney and rose to prominence to challenge for the mayorality of New York City in 1989. Though he lost that year to David Dinkins he went on to win in 1993, again in 1997 and stayed mayor until 2001. He was known for his tough stance on crime, his efforts after 9/11 and is responsible for many forward reforms in the city. In 2000, he ran for senate against Hilary Clinton before being diagnosed with cancer. At the same time he was involved in an extramarital affair with Judith Nathan. He then decided to run for president in 2007. Identifying the New York Times newspaper archive\cite{nytarchive} as the best source of primary material, the historian formulates her intent of finding information related to Giuliani with the keywords \texttt{rudolph giuliani} and sets the publication date filter to 1987 -- 2007 only to get the following results:

\begin{table}[ht]
\centering
  \footnotesize
   \scalebox{0.7}{
  \begin{tabular}{@{}ccl@{}}\toprule

    Rank & Year & Headline \\
	 \midrule
	 1 & 2007 & \texttt{In His Own Words}\\

	2 & 2007& \texttt{Giuliani Is Expected to Sell One of His Three Businesses} \\

	3 & 2007&\texttt{Giuliani Is Selling Investment Firm}\\

	4 & 2007&\texttt{'08 Candidacy Could Shake Up Giuliani's Firm}\\

	5 & 2006&\texttt{Giuliani Building Network of Donors, a Backer Says}\\

    \bottomrule
  \end{tabular}}
\end{table}

None of the result documents mention his stance on \textsf{crime}, \textsf{cancer}, \textsf{Hilary Clinton}, \textsf{World Trade Center}, \textsf{David Dinkins} or his reforms. Arguably a better set of results for her to scope out her topic is a diversified set of top documents covering important
aspects which could be entities like Hilary Clinton and David Dinkins but also from important time points so that she can contextualize or reformulate her query.

Learning from the experiences of our colleagues at the British Library and the Institute of Historical Research in London (cf. Section~\ref{sec:hist}), in this work, we propose a novel document retrieval task which intends to present the most relevant results from a topic-temporal space. This problem can be seen as a generalization of the classical diversity problem by adding a temporal dimension. The topical diversity focuses on presenting results from different subtopics while the temporal diversity ensures that the documents returned are primary in nature. However, the challenge in adapting existing diversity-based approaches is the following. Firstly, traditional diversity approaches~\cite{carbonell1998use,santos2010exploiting,agrawal_diversifying_2009,Carterette:2009:PMR:1645953.1646116,dang_diversity_2012} tend to diversify typically on the topical aspect and do not take time into account. As a result, documents retrieved might still cover a good number of aspects but (a) might be from the same time period disregarding the temporal salience of the aspect and (b) might return secondary sources even when more relevant primary sources are present. For the query \texttt{rudolph giuliani}, such methods would not guarantee that documents are from the important time periods like '89, '93, '00, '01 and '07.

Time-aware approaches which take into account latent topics or aspects like~\cite{ecir/NguyenK14} are optimized to present results which are valid at querying time or in other words reward recency. On the other hand \cite{lm+t+d} diversifies based on time without explicitly considering topical aspects. Although this ensures that results are temporally distant from each other, as a consequence of the inherent topic-agnostic nature they still might belong to similar aspects. In our example, if the historian uses a temporal diversification retrieval model there is no guarantee that results returned from 2000 and 2001 will definitely cover the WTC, cancer, Hilary Clinton and Judith Nathan. Finally, multi-dimensional approaches to diversity treat both time and aspects similarly which is not always desirable since both these dimensions have different semantics. We on the contrary, explicitly model both the topical and temporal aspects of a document by treating time as a first-class citizen in our model. In our approach, called \textsc{HistDiv}, the temporal space is based on primary (publication times) and secondary sources (temporal references in text) and the topic space is based on the entities present in the news article. We then jointly diversify both in the aspect and time dimensions discounting each of these dimensions based on semantics unique to each. In sum we make the following contributions:

\begin{itemize}
 	\item We introduce the notion of \emph{Historical Query Intents} and model this as a search result diversification task on both the aspect and time dimensions for historical search.

 	\item We develop a novel retrieval algorithm called \textsc{HistDiv} which jointly diversifies both dimensions by appropriately discounting the contribution of aspects and time. 

 	\item We establish the effectiveness of our method by building a test collection based on the 20 years of the \emph{New York Times Collection} as a dataset and a workload of 30 manually judged queries which will be made available to the community. The quantitative results show we outperform our competitors in subtopic recall at the cost of precision.  
 	
 	\item Finally, we conducted a qualitative study with the target users to confirm if this loss in precision truly harms the quality of the overview derived from the ranking.
 \end{itemize} 

\section{How do historians search?}
\label{sec:hist}

A historian's corpus of study is often an archive. Archives consist of
time annotated records, from the distant and recent past, categorized
as primary and secondary sources. 


\textbf{Need for Overview and Context:} A vital step in a historians
search process is to browse the archive in order to get an
\emph{overview} of material available on a topic. This allows them to
identify potential areas of interest which subsequently lead to more
focused queries in the next search phase. Undoubtedly, the onset of
digital archives has greatly improved information access but a key
requirement of historians still is to obtain an overview of multiple
aspects of the topic~\cite{smith2004historians}. This overview allows
them to not only find more specific topics of study but also to
contextualize their results by studying the temporal and spatial
vicinity of the topic.

\textbf{Focus on primary sources:} The main preoccupation of
historians is the reading of primary sources while secondary sources
are useful when no primary evidence exists and for tracing references
to cited primary sources \cite{case1991collection}. Even with the
onset of digital archives, \cite{smith2004historians}
re-enforces the historian's preference and focus on primary sources.

\textbf{Historian Search Behavior:} Based on existing literature and a short survey 
in collaboration with the British Library (BL) and the Institute of
Historical Research in London, we found that the current way of searching
digital archives is a two-stage iterative querying process. In the first
stage, keyword queries on broader topics are issued and further
reformulated by the use of a combination of filters (time, source,
entity, etc.) and facets to gain an overview of the results. Subsequently, more
specific queries on each aspect are prepared to serve their
information need. \cite{bingham2010digitization} finds keyword
search (labeled a \emph{blunt instrument}) in archives to be most
effective when the user is very precise and focused in his
search. Hence, the initial stage is indeed the most cumbersome,
firstly because it leads to a large number of results which then have
to be read in its entirety, and secondly because of the increased usage of
filters. Examples of such general queries are queries about entities,
recurring events, conflicts etc.  



In general, historians working with physical archives are privy to themed sub-collections
created by archivists. An entity search in these smaller collections
may produce plenty of results but nearly all tend to be relevant. On
the other hand, a news archive is a single large collection only
subdivided based on time, hence searching for entities, especially
popular ones, can return many records.

As we will show in the remainder of this
paper, the joint diversification of aspects and time can lead to
better ranking of documents for the initial search phase, involving these broad keyword queries, in newspaper archives. 
\section{Historical Search Task}
\label{sec:hst}

\begin{figure*}[t!]
\centering
  \fbox{\includegraphics[width=0.8\textwidth]{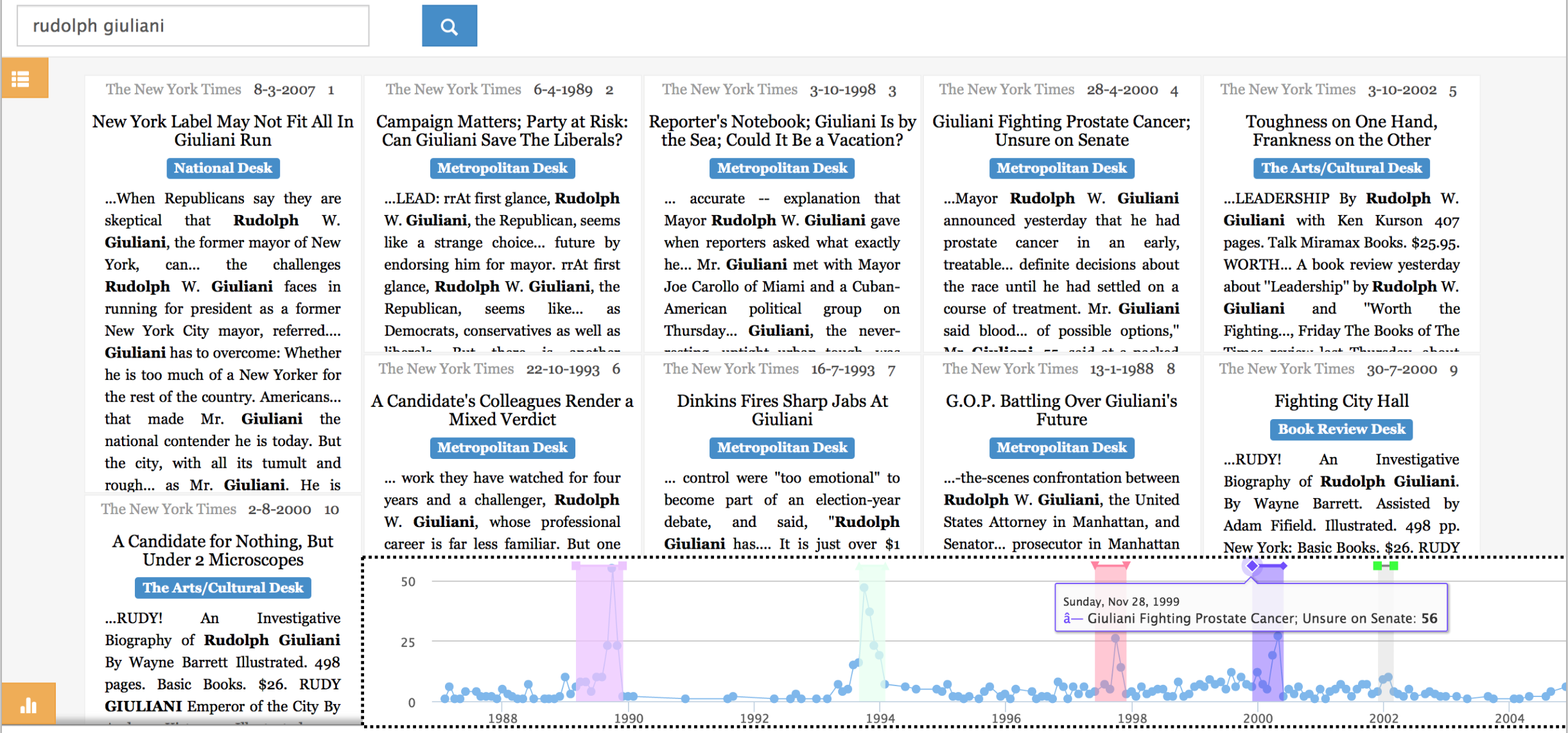}}

\caption{\small{The \textsc{HistDiv} search system. The search results are layed out to mimic a newspaper. The coloured areas in the timeline represent bursts detected from the temporal distribution. Results shown are for the top 10 out of 26,000 results returned for the query Rudolph Giuliani.}}

\label{fig:example}
\end{figure*}

A \emph{Historical Query Intent} is the moniker we choose to describe a user's intent to cover as many historically relevant subtopics and time windows for a given topic. According to \cite{jones_temporal_2007}, a temporal query intent is used to specify queries which are either atemporal, temporally ambiguous or unambiguous. We however deal with a special case of temporally ambiguous queries which have an explicit information need for the past. Additionally, historical query intents also deal with ambiguity with respect to the aspect and time dimensions. In this section we formally present our input model and define our historical search task problem based on historical query intents, and discuss how to measure the effectiveness of proposed approaches. 

\subsection{Model} 
\label{sub:model}


\textbf{Document Model:} We operate on a document collection $\mathcal{D}$ where each document $d_p \in \mathcal{D}$ has a publication time point $p$. The content of $d_p$, for instance "\underline{Dinkins} pulls negative ad about \underline{Giuliani} as the race for Mayor draws closer", can be represented by a set of aspects such as \textsf{\{David Dinkins, Rudolph Giuliani\}}. The set of aspects describing the content of $d_p$ is denoted by the set $ A(d_p) = \{a_1,\dots a_n\}$ where $a_i$ is a single aspect like \textsf{David Dinkins}. The document can also contain temporal references such as "last year" or "2001" which can be useful indicators of important time intervals. The set of temporal expressions contained in $d_p$ is given by $E(d_p) = \{I_1, I_2, \dots I_n\}$ where $I$ is an arbitrary time interval with a definite begin and end timestamp.

\textbf{Temporal Model:} We adopt a discrete notion of time for the collection and assume that a time-stamp $t_i$ is a positive integer and is computed periodically, with a fixed granularity $\Delta$, from a reference point in the past $t_0$. The discretized time span of the collection is denoted by an ordered set $W = \langle t_0, t_1, \dots t_n \rangle$ such that $\Delta = t_{i+1} - t_i$. We define $\delta_i= [ t_i, t_{i+1} )$ as the elementary time interval between two consecutive time points. A document $d_p$ is published in the interval $\delta_i$ if $t_i \leq p < t_{i+1}$ and is given by the function $\Lambda(p) = \delta_i$. The set of all elementary time intervals of size $\Delta$ in $W$ is called the temporal space
\vspace{5pt}
 $$\mathcal{T} = \{ \delta_i \,\,| \,\, \delta_i \in \bigcup_{d_p \in \mathcal{D}} \Lambda(p) \}$$ 
\\


\textbf{Aspect-Time Space:} For a given query $q$, $\mathcal{R}_q$ is set of top-K documents retrieved such that $\mathcal{R}_q \subseteq \mathcal{D}$. The time space relevant to $q$ is the set 
\vspace{5pt}
\\
$$\mathcal{T}_q =\{ \delta_i \,\,| \,\,\delta_i \in \bigcup_{d_p \in \mathcal{R}_q} \Lambda(p) \}$$
\\
\vspace{5pt}
such that $\mathcal{T}_q \subseteq \mathcal{T}$. Similarly we define the aspect-space of $q$ as the set of all aspects found in documents from $\mathcal{R}_q$ denoted by 
\vspace{5pt}
\\
$$\mathcal{A}_q = \{ a_i \,\,|\,\, a_i \in \bigcup_{d_p \in \mathcal{R}_q} A(d_p)\}$$
\\
\vspace{5pt}
where $\mathcal{A}_q \subseteq \mathcal{A}$. However, not all aspects are relevant in all time intervals. For example, the aspect \textsf{World Trade Center} for the query \texttt{rudolph giuliani} is historically irrelevant for time intervals before 2001. To this end, we define a combined aspect-time space 
\vspace{5pt}
\\
$$\mathcal{AT}_q \subseteq \mathcal{A}_q \times \mathcal{T}_q$$
\\
\vspace{5pt}
 which contains aspect-time pairs encoding subspaces which are both temporally and aspect-wise relevant. 
\\
\vspace{5pt}
$$
	\mathcal{AT}_q = \{ (a_i, \delta_j) \,\,\,|\,\,\, a_i \in \mathcal{A}_q \wedge \delta_j \in \mathcal{T}_q \}.
$$
\\
\vspace{5pt}
A result document $d_p$ for query $q$ with aspects $A(d_p)$ and temporal expressions $E(d_p)$ is said to be relevant to an aspect-time pair $(a_i, \delta_j) \in \mathcal{AT}_q$ if $a_i \in A(d_p)$ and $\Lambda(p) = \delta_j$. 

\subsection{Problem Definition}

The historical search result diversification problem or simply the \emph{historical search task} intends to find a re-ranking $S$ of an initial result set $\mathcal{R}_q$ that has maximum coverage and minimum redundancy with respect to different aspect-times underlying $q$. In other words, it is the standard search result diversification task but over the $\mathcal{AT}_q$-space which encodes the historical query intent of obtaining relevant documents from the most important aspects from the time-period/s it is important in. As shown by~\cite{agrawal_diversifying_2009}, search result diversification is a bi-criterion optimization problem which can be reduced from the maximum $k$-coverage problem and hence is $\mathcal{NP}-$hard.

\subsection{Evaluation Measures}

Given that we define a new two-dimensional solution space, we could re-use the standard diversity-based retrieval measures for evaluating approaches to the historical search task, considered on the $\mathcal{AT}_q$-space. Subtopic-recall $\textsf{SBR}_{q,k}$, for instance, for $\mathcal{R}_q$ at depth $k$, can be computed as: 
$$\frac{ \left| \,\,\bigcup_{d_p \in \mathcal{R}^k_q} \{ \, (a_i, \delta_j) \,\,\,| \,\,\,  a_i \in A(d_p) \, \wedge \Lambda(p) = \delta_j  \} \right| }{ \left| \mathcal{AT}_q\right| } .$$

Similarly for other diversity metrics suggested in \cite{clarke2008novelty} like intent aware ERR (\textsf{IA-ERR}), intent aware precision (\textsf{IA-P}), mean average precision \textsf{MAP} and $\alpha$-NDCG (\textsf{NDCG}) we can substitute the subtopic space with the $\mathcal{AT}_q$-space.

\section{Related Work}
\label{sec:rel-work}

Before we describe the \textsc{HistDiv} approach for historical search tasks, in this section we outline the relevant existing literature. Our problem has overlap with areas relating to temporal representation and temporal retrieval models under temporal IR, and with works on search result diversification. 

\vspace{-2mm}
\subsection{Temporal Information Retrieval} 

Temporal IR has emerged as an important subfield in
IR with the goal to improve search effectiveness by exploiting
temporal information in documents and
queries~\cite{campos_survey_2014}. The value of the temporal dimension
was clearly identified in~\cite{Alonso:2007} and has
led to a plethora of work which utilizes temporal features in query
understanding~\cite{jones_temporal_2007,Metzler:2009}, retrieval
models~\cite{berberich_language_2010,Dong:2010a,Dong:2010b,brucato_metric_2014}, temporal
indexing~\cite{berberich_time_2007,anand_index_2012},
clustering~\cite{Alonso:2009} and query
modelling~\cite{radinsky_behavioral_2013,peetz2012adaptive,choi2012temporal}. A
survey by Campos et. al.~\cite{campos_survey_2014} gives an elaborate
overview of the field.

\textbf{Improving Ranking using temporal features} One of the first
algorithms to incorporate time in search result ranking was suggested
in \cite{li2003time}. They used a temporal language model
approach where time and term importance are handled
implicitly. Various approaches have been suggested that consider time
more explicitly. \cite{berberich_language_2010} proposes a language modelling approach taking into account the temporal expressions in the query and document text. \cite{brucato_metric_2014} on the other hand, taking a non-probabilistic interpretation of relevance, defines \emph{temporal scope similarity} between queries and documents in \emph{metric spaces}. In both these works, similarity between the temporal references in the query and documents are used to rank documents. In our query model, we never make any assumptions on the presence of temporal references nor do we model the similarity of query and document based on temporal references. Another line of work in this domain considers the freshness or recency of a document when ranking \cite{Dong:2010a}.


\textbf{Finding important time periods} An important ingredient in our
retrieval model is finding temporal priors for different time
points. \cite{jones_temporal_2007} estimates a
probability distribution over different time points for each query
called the temporal query profiles using publication times of the
documents. In \cite{Setty:2010}, the authors exploit the
publication dates to identify important time points for a given query
by contrasting rankings for adjacent time points. However, neither of
them utilize secondary sources or temporal references in
text. Recently, \cite{Gupta:2014} suggests using temporal
references for ranking time intervals for a given temporal query. In
our work, we use both publication dates (primary sources) and temporal
references (secondary sources) to assign temporal priors to each granular time
 window akin to a temporal profile~\cite{jones_temporal_2007}.


\subsection{Search Result Diversification} Diversity in search (both
explicit and implicit) has seen a rich body of literature lately
in~\cite{dang_term_2013,carbonell1998use,agrawal_diversifying_2009,santos2010exploiting,Carterette:2009:PMR:1645953.1646116,zhu_learning_2014,liang_fusion_2014,dang_diversity_2012}. Search result diversification aims to maximise the overall relevance of a document ranking to multiple query aspects, while minimising its redundancy with respect to these aspects. Existing approaches differ in the way they model different query aspects. Implicit approaches, like~\cite{carbonell1998use}, assume similar documents cover similar aspects and do not model aspects. Explicit approaches model aspects in a variety of ways using query logs, taxonomies etc. We also model aspects explicitly by using entities found in text documents as their aspects. Also, none of the previous approaches take time into account or model the historical information intent.

\textbf{Temporal diversification}  \cite{lm+t+d} proposes a diversification model which considers
time windows as a set of intents for a query while modeling the
importance of each intent as the weight of its burst. In traditional
aspect-based diversification tasks
like~\cite{clarke2009overview,clarke2011nist}, intent importance is
considered static over time. However, intent importance was shown to
vary across time; thus affecting the diversity evaluation of queries
issued at different time points \cite{zhou2013impact}. Keeping this in
mind, \cite{ecir/NguyenK14} considers the time at which
the query is issued to diversify intents based on their temporal
significance at that time. Their approach also explicitly models time
and aspects, although latent, but rewards recency. \textsc{HistDiv},
on the other hand, is query-time agnostic, since it is intended for
historical search, and seeks to diversify documents based on both time
and aspects.

\section{The HistDiv Approach}
\label{sec:approach}

\subsection{Approach Overview}
\label{sec:overview}

The challenge in designing a retrieval model for the historical search task is to identify important time intervals and aspects. More importantly, the documents which optimize both dimensions and are relevant to the user intent. Towards this, we first model the temporal space for the initial result set $\mathcal{R}_q$. We build a probability distribution $P(\delta_i|q)$ for $\delta_i \in \mathcal{T}_q$ over the entire time span $W$ taking into account the publication times and temporal references (mined from document text). Such a temporal profile, as shown in Figure~\ref{fig:example} helps us isolate the important time-periods in the result timeline (described in Section \ref{sub:temporal_priors}). Next in Section \ref{sub:aspect_priors}, we detail how we build priors for the aspects of the documents. Finally in Section \ref{sub:approach}, we present our diversification algorithm \textsc{HistDiv} which takes into account textual relevance, temporal sensitivity and aspect importance along with the \emph{typical semantics} of the temporal and aspect domains to maximize coverage in the $\mathcal{AT}_q$ space.

\subsection{Building Temporal Priors}
\label{sub:temporal_priors}
To find important time intervals, we build a probability distribution $P(\delta_i|q)$ over the entire time span by projecting both the publication times and the reference times into $\mathcal{T}_q$. The temporal references are treated as secondary sources and can be used as indicators of relevant primary sources. For computing the distributions we use the document counts published in a time interval $\delta_i$ (contrary to using top-k relevance scores for profile generation~\cite{jones_temporal_2007}). The probability $P_{pub}(\delta_i | q)$ is the fraction of all documents in $\mathcal{R}_q$ published in $\delta_i$. To compute $P_{ref}(\delta_i | q)$ we first estimate the contribution of an interval $I \in E(d_p)$ as $\frac{1}{|I|}$ for all constituent time intervals $\delta_i$. Finally, we employ a language modeling strategy to smooth the probability distribution of the publication times $P_{pub}$ with the background distribution of the temporal references $P_{ref}$ with a mixing parameter $\theta$ to arrive at a distribution $P(\delta_i | q)$ :
$$
	P(\delta_i | q) \,=\, \theta.P_{pub}(\delta_i | q) \,+\,(1-\theta).P_{ref}(\delta_i | q).
$$

In our experiments we demonstrate the value of estimating the time prior by comparing it to a temporal diversification baseline that assumes equal distribution of $\delta_i$ called \textsc{EqT}.

\subsection{Aspect Modeling}
\label{sub:aspect_priors}

Historians are particularly interested in events which can be described using groups of entities associated with specific time intervals. Keeping this in mind, we use entities mentioned in the document text as our aspects.Traditional aspect-space diversification methods, like \textsc{Ia-Select}~\cite{agrawal_diversifying_2009} and \textsc{Pm2}~\cite{dang_diversity_2012}, estimate the probability $P(a_i|q)$ assuming the collection is static. For historical search, time is an essential factor and needs to be considered when estimating  $P(a_i|q)$. Consider a document $d_p$ published in time interval $\Lambda(p)=\delta_j$;  an aspect $a_i \in A(d_p)$ can be temporally diverse if it occurs in documents from different time intervals. Hence the aspect $a_i$ has a probability distribution across time intervals $\delta_j \in  \mathcal{T}_q$. Consequently, the prior probability of an aspect $a_i \in \mathcal{A}_q$ in a time interval $\delta_j \in \mathcal{T}_q$ is given by:
$$
	P(a_i|q,\delta_j) = \frac{\big| \{d_{a_i,p} \, \, \, | \, \, \, a_i \in A(d_p) \wedge \Lambda(p) = \delta_j\} \big|}{\left|\{ d_p \, \, | \, \, \Lambda(p) = \delta_j \} \right|} 
$$
where $d_{a_i,p}$ is a document tagged with aspect $a_i$ published in the interval $\delta_j$. Notice in the Figure~\ref{fig:example} that the event \textsf{mayoral campaigns} is recurring every 4 years. Hence the aspects representing \textsf{mayoral campaigns} will have a higher $P(a_i|q,\delta_j)$ in certain time intervals $\{ \delta_{1989}, \delta_{1993}, \delta_{1997}\}$ when compared to the others.

\begin{algorithm}[!t]
  \small
  \SetKwFunction{Decompose}{decompose}

  \KwIn{ $k, q, \mathcal{A}_q, \mathcal{R}_q, \mathcal{T}_q, V(d|q), S = \emptyset$}
  \KwOut{Set S of diversified documents}

  \ShowLn$\forall a \in \mathcal{A}_q \: , \: \forall \delta_i \in \mathcal{T}_q, \:\: U_{aspect}(a|q, S, \delta_i) =$ refer Eqn. 1

  \ShowLn$\forall \delta_i \in \mathcal{T}_q, \:\: U_{time}(\delta_i|q,S) = $ refer Eqn. 2

  \ShowLn\While{$|S| \leq k$}
  {
     
    \ShowLn\While{$d \in R$}
    {
      \ShowLn $g(d|q, S) \leftarrow  \alpha . V(d|q) \,\,\,+\, (1-\alpha).( \beta . \sum_{a}^{A(d)} U_{aspect}\,\, + \,\,(1-\beta). U_{time})$
    }

    \ShowLn $d^{*} \leftarrow argmax_{d} \:\: g(d|q, S)$

    \ShowLn $S \leftarrow S \,\,\cup \,\, \{d^{*}\}$
  }

  \ShowLn\Return{$S$}

  \BlankLine

  \caption{The \textsc{HistDiv} Algorithm}
  \vspace{-2mm}
  \label{alg:ATD}
\end{algorithm}
\subsection{The \textsc{HistDiv} Algorithm}
\label{sub:approach}

In classical diversification approaches like~\cite{agrawal_diversifying_2009,dang_diversity_2012,mdiv}, each document is assigned a utility score computed using textual relevance (denoted as $V(d|q)$ in Algorithm \ref{alg:ATD}) and aspect importance. Most approaches employ a greedy algorithm which selects the candidate documents that maximize utility with respect to the uncovered aspects in each iteration. \textsc{HistDiv} considers both aspect and time dimensions (with their special semantics) and operates in a similar manner treating both the topical and temporal aspects as sets thereby retaining the $(1-1/e)$ approximation guarantee. However, we differ significantly from previous approaches in the way we interpret and compute the utility of each dimension. Traditional diversification algorithms model only aspects and maximize coverage in the space $\mathcal{A}_q$. Since the objective is to maximize coverage in the $\mathcal{AT}_q$ space we first consider how they can be adapted to model the required space $\mathcal{AT}_q$. 

\textbf{Temporally augmented aspect space:} A na\"ive approach to introduce time could be to enrich the aspect space by adding time intervals as new aspects. For instance for $d_p$ published in $\delta_i$ we can add $\delta_i$ to $A(d_p)$. In our experiments we use this method to create two variations of \textsc{Ia-Select} and \textsc{Pm2} called \textsc{E-Ia-Select} and \textsc{E-Pm2}.

\textbf{Linearizing aspect space with time:} Since we deal with two dimensions we can project or \emph{linearize} the temporal dimension onto the aspect dimension. More formally, the result of linearization is the set of $m$ aspects $\{ a_1\delta_i, \ldots, a_m\delta_i \}$ for document $d_p \in \delta_i$ which is used to alter \textsc{Ia-Select} and \textsc{Pm2}. We name these two variations \textsc{T-Ia-Select} and \textsc{T-Pm2} and also use them as baselines in our experiments.

An alternative would be to keep the dimensions separate like the multi-dimensional approach proposed in~\cite{mdiv} (referred to as \textsc{MDiv} henceforth). In this general framework, for the diversification of $n$ arbitrary dimensions, the utility score $g(d|q, S)$ computation reflects how the dimensions are combined. The marginal utility of aspects given a document $d$ is computed based on rank of $d$ for the given aspect $a_i$. We can naturally add time as a second dimension and use it for diversification. We use \textsc{Mdiv} as a competitor in our experiments.

A key drawback of both these approaches is that they do not consider the fact that: (a) temporal aspects are ordered and thus have special semantics (b) temporal and topical aspects are interrelated. Hence \textsc{Mdiv}'s assumption of dimensional independence and identical discounting function for both dimensions might not yield optimal results. A retrieval model designed for a historical search task however should take this into account while computing the utility of aspects and time intervals to optimize coverage of the $\mathcal{AT}_q$ space (see Section~\ref{sec:hst}). 

In \textsc{HistDiv}, the utility of a document in the aspect space is measured by $U_{aspect}(a_i|q, S, \delta_j)$ with the exception that we treat an aspect in various time windows differently. We discount aspects in a neighborhood defined by the $w$ so that $P(a_i|q,\delta_j)$ is strongly discounted if $\delta_j$ is temporally closer to a document $d_p \in S$ and $a_i \in A(d_p)$. We use the decay function suggested in \textsc{OnlyTime}~\cite{lm+t+d} to discount aspects across time. In this way, we avoid the time agnostic property of standard topical diversification retrieval models which may select the right aspects but will discount the aspect for the entire span of the collection thereby reducing the probability of selecting documents for the same aspect from other important time intervals. The utility of an aspect, $U_{aspect}(a_i|q, S, \delta_j)$, is
$$
P(a_i|q,\delta_j) \prod_{d_p \in S} \left ( 1 - \frac{1}{1+e^{-w+|t_j-p|}} \right ) \, \,  \,  \, \,(1)
$$

$t_j$ denotes the boundary time point of $\delta_j$. Like \textsc{OnlyTime} we can set $w$ to the size of $\Delta$. 
The limitation of modeling utility this way is that we are restricted by the fixed $w$. Consider bursts, where a high number of documents from multiple consecutive $\delta$ typically discuss a single event (especially if $\Delta$ is small). If we select a document about this event from the edge of the burst then we face two potential issues: (i) we may assign high utility to documents which are about similar aspects from two temporally distant intervals but still refer to the same event (ii) we heavily discount documents just outside the burst (temporally close but sufficiently different) unfavorably. Both of these issues lead to a potential drop in subtopic recall which we address by using the burst detection technique suggested in \cite{peetz2014using}. To detect the set of bursts $B_q$, this technique utilizes the mean and standard deviation of a fixed width sliding window across $W$. 

We can now vary $w$ depending on the position of $d_p$ within its corresponding burst ($b_i \in B_q$) or non-bursty interval ($b_i \in \hat{B_q}$) and use $U_{aspect}(a_i|q, S, \delta_j)$ as before. $w$ is then computed as follows: 
\[ w = \left\{
    \begin{array}{lr}
      |p-begin(b_i)| & : p \ge t_j\\
      |p-end(b_i)| & : p < t_j
    \end{array}
  \right.
  \]
In the time dimension, we need to be wary of discrediting a time interval too heavily. \textsc{OnlyTime} produces a result set with high temporal diversity by selecting relevant documents from important intervals and discounts those intervals heavily with the aforementioned decay function. This approach to discounting bursts doesn't consider the fact that a single burst could consist of many diverse aspects. For example, in 2000-2001 Giuliani was divorced, diagnosed with cancer and was involved in helping New York recover from 9/11. Hence unlike \textsc{OnlyTime}, we discount the interval in the time dimension of $d_p \in S$ by the weighted proportion of aspects covered by it. The utility of time, $U_{time}(\delta_i|q,S)$, is 
$$
  P(\delta_i|q)  \prod_{d_p \in S} \left ( 1 - \frac{|\,( d_{a_j,p^*} \,|\,a_j \in A(d_p) \wedge \Lambda(p^*) = \delta_i)|}{| (d_{p^*} \, | \,  \Lambda(p^*) = \delta_i )|} \right ) \, \, \,(2)
$$

With burst detection, for all $d_p \in S$ we simply discount all time intervals $\delta_i$ contained in it's corresponding bursty / non-bursty interval.

The essence of our approach lies in our temporal interpretation of aspect utility and aspect aware interpretation of time utility. This interpretation helps us maximize coverage in the joint space $\mathcal{AT}$, as shown in our experiments, when compared to pure aspect based diversification, pure time based diversification and multidimension diversification. Algorithm~\ref{alg:ATD} shows the iterative process in which documents are selected in to result based on the utility $g$, where aspect and temporal utilities are traded-off by the parameter $\beta$. The parameter $\alpha$ trades-off the impact of document relevance $V(d|q)$ with the utility of the two dimensions.

\section{Test Collection } 
\label{sec:test-collection}

\todo{
	\begin{enumerate}
	  \item How exactly were subtopics found? (with the help of historians)
	  \item collection will be available to the public
	  \item Collection characterisitics
	\item Search Topics characterisitics
		\item gathering judgements
	\item Measures ??
	\end{enumerate}
}
In this section we detail the test collection which we constructed to evaluate our approach. There exist well known collections for the standard diversification task (diversity tasks in the TREC Web track), temporal information retrieval (Temporalia'14~\cite{joho2014ntcir}) as well as web archives~\cite{costa2012evaluating}; however to the best of our knowledge there are no established test collections to measure the effectiveness of retrieval models designed for diversification of the $\mathcal{AT}_q$ space in news archives. Hence we choose to build our own collection guided by the target user group - historians, whose judgments will be made available to the research community.

\textbf{Metrics:} The evaluation metric for a search task should reflect the user's goal. For users with historical query intents the objective is to find primary sources of information from all the important aspects and time periods. The primary metric we choose to measure the effectiveness of a retrieval model is \textsc{SBR} (subtopic recall) in the joint aspect-temporal space $\mathcal{AT}_q$ because of the recall oriented nature of historical query intents.   

\textbf{Document Corpus:} As a corpus we use the \emph{Annotated New York Times} collection~\cite{nytarchive} which qualifies as a suitable news archive since it spans for 20 years, i.e., 1987 - 2007. Although there exists larger news corpora, they span for a shorter time duration reducing the likelihood of having ample primary sources .Also, the timestamps associated with the articles are accurate and do not have to be estimated as in other web collections. The corpus consists of 1.8 million articles from all sections of the newspaper including the editorial desk, arts, technology and literature making it replete with various aspects interesting for historians.


\todo{
  \begin{enumerate}
    \item Tie it back to section 2 about the characteristics of historical search
  \end{enumerate} 
}

\textbf{Search topics:} Topics, with a historical intent, for our test collections are derived from experts who held discussions with historians at the Institute of Historical Research as well as from insights in Section \ref{sec:hist}. These experts first described the intents verbosely and then proceeded to identify keywords that represent it. Since our corpus is a newspaper daily for the USA, topics are chosen from a set of historically relevant issues related mostly to the USA and a few from more global issues. To define the \emph{subtopics} of each topic the experts were guided by the history sections from the relevant Wikipedia articles. To confirm or modify subtopics, they explored the corpus with a simple keyword search interface whenever necessary. The chosen subtopics are also qualified by a set of relevant time periods. The experts chose time periods of relevance to the subtopic by consulting relevant Wikipedia articles and defined the interval size as the period in which they found primary sources in the corpus. Depending on the type of subtopic the time interval can span months (Giuliani's efforts in the \textsf{aftermath of 9/11}) or years (Giuliani's \textsf{senate run}). Each subtopic can also have multiple time intervals like Giuliani's \textsf{mayoral election campaigns} which are relevant during \texttt{1989},\texttt{1993} and \texttt{1997}. Time intervals can also overlap each other, for instance Giulaini's \textsf{personal life} (struggle with cancer) and his \textsf{senate run}.


\lstset{
  language=xml,
  tabsize=1,
  caption=Excerpt of a topic in the workload,
  label=xml:topic,
  rulesepcolor=\color{gray},
  xleftmargin=0pt,
  framexleftmargin=0pt,
  basicstyle=\footnotesize,
  keywordstyle=\color{blue}\bf,
  commentstyle=\color{OliveGreen},
  stringstyle=\color{red},
  breaklines=true,
  basicstyle=\tiny,
  emph={topic,subtopic,time,query},emphstyle={\color{magenta}}}
\begin{lstlisting}
<topic>
	<query>rudolph giuliani</query>
	<desc>I want to know the history of Rudolph Giuliani</desc>
	<subtopics>
		<subtopic>
			<desc>Mayoral campaigns</desc>
			<time>[{01.01.1989 - 31.12.1989}, {01.01.1993 - 31.12.1993}, {01.01.97 - 31.12.1997}]</time>
		</subtopic>
		<subtopic>
			<desc>Senate race</desc>
			<time>[{01.01.2000 - 31.12.2000}]</time>
		</subtopic>
		<subtopic>
			<desc>Efforts after 9/11</desc>
			<time>[{11.09.2001 - 01.04.2002}]</time>
		</subtopic>
		.
		.
		.
		.
	</subtopics>
</topic>
\end{lstlisting}

We have a total query workload of 30 topics. On average there are 5 subtopics per topic and each subtopic has at least one relevant time interval. The types of topic chosen are inspired by the characteristics defined in Section \ref{sec:hist}, i.e, broad topics related to entities like \texttt{Rudolph Giuliani} and the \texttt{Atlantic City}, major events like the \texttt{reunification of germany} and \texttt{team usa soccer world cup} as well as controversial subjects like \texttt{gay} \\ \texttt{marriage} and \texttt{sarin gas}. A key assumption made when creating subtopics is the omission of historical facts that lie outside of the 20 year time period of the NYT corpus.

\todo{
	\begin{enumerate}
		\item list of queries from NYT portal (24.01.15) -modern love, obama, charlie hebdo, india, state of the union
	\end{enumerate}
}

\textbf{Pooling:} We devise suitable baselines (detailed in Section~\ref{sub:competitors}) and "submit" runs for each baseline corresponding to all possible parameter settings. By doing so we increase the coverage of documents for each topic and improve the diversity of the pool. We chose a run size of top 20 for all topics and the pool size was set to 300 documents per topic. We overall generated nine competitors which produced on average 20 runs per baseline. To gather relevance judgments we use the Cranfield paradigm~\cite{voorhees2002philosophy}. Trained evaluators were instructed to assign binary relevance judgments to topic, subtopic, document triples. Each document was judged once with high confidence. Once the pools were evaluated, a standard robustness test was carried out with $\textsf{SBR}$ over $\mathcal{AT}_q$ as the primary measure. We selected 25\% of the query workload at random and split them into two equal sets. We selected 50\% of the runs at random for retrieval depth 10 and ranked the system runs for both sets of queries by $\textsf{SBR}$. We found that the rankings were consistent (high Kendal's tau) for p<=0.05.

\section{Experimental Evaluation}
\label{sec:experiments}

Before we present our results, we detail our experimental setup in which we describe how we mine our aspects and select baselines. Then to assess the effectiveness of our approach we first present in Section~\ref{sub:subtopic_recall} the overall retrieval effectiveness across different retrieval depths, assess the impact of varying the granularity $\Delta$ and also highlight certain drawbacks. Next, in Section~\ref{sec:user-centric-study} we discuss interesting insights from a user study to estimate the quality of an overview produced by different rankings and finally summarize take-aways from our experiments.


\begin{table*}[ht!]
\centering 
\footnotesize 
\small
\begin{tabular}{
@{}lrrlcrrlcrrl@{}}\toprule
& \multicolumn{3}{c}{\textbf{k=10}} & \phantom{abc} & \multicolumn{3}{c}{\textbf{k=15}} & \phantom{abc} & \multicolumn{3}{c}{\textbf{k=20}}\\ 
\cmidrule{2-4} \cmidrule{6-8} \cmidrule{10-12}
& \multicolumn{1}{c}{$A$} & \multicolumn{1}{c}{$T$} & \multicolumn{1}{c}{$AT\,(W/L\%)$} && \multicolumn{1}{c}{$A$} & \multicolumn{1}{c}{$T$} & \multicolumn{1}{c}{$AT\,(W/L\%)$} &&  \multicolumn{1}{c}{$A$} & \multicolumn{1}{c}{$T$} & \multicolumn{1}{c}{$AT\,(W/L\%)$} \\ \midrule
\textsc{Lm} & 0.706 & 0.060 & 0.428 && 0.752 & 0.085 & 0.491 && 0.780 & 0.091 & 0.518\\[1pt]
\textsc{Ia-Select}$^{\circ}$& 0.722 & 0.039 & 0.442 (23/23) && 0.766 & 0.047 & 0.491 (20/26) && 0.841 & 0.055 & 0.516 (20/23)\\[1pt]
\textsc{Pm2}$^{\star}$ & 0.707 & 0.069 & 0.429 (16/20) && 0.794 & 0.082 & 0.471 (10/23) && 0.817 & 0.097 & 0.509 (16/26)\\ \midrule

\textsc{Tia-Select}$^{\bullet}$ & 0.614 & 0.039 & 0.380(23/36) && 0.717 & 0.047 & 0.433 (20/43) && 0.770 & 0.055& 0.470 (20/26)\\[1pt]
\textsc{T-Pm2}$^{'}$ & 0.551 & \textbf{0.088} & 0.308 (13/50) && 0.680 & 0.106 & 0.408(20/43)&& 0.761& 0.128 & 0.453 (16/33) \\[1pt]

\textsc{E-ia-Select}$^{\ddagger}$ & 0.700 & 0.062 & 0.435 (23/23) && 0.776 & 0.084 & 0.501 (23/23)  && 0.837 & 0.095 & 0.524 (23/20) \\[1pt]
\textsc{E-Pm2}$^{\dagger}$ & 0.692 & 0.061 & 0.422 (6/16)  && 0.766 & 0.083 & 0.469 (6/26) && 0.816  & 0.098 & 0.495 (10/26) \\[1pt]

\textsc{EqT}& 0.714 & 0.076 & 0.440 (16/13) && 0.766 & 0.097 & 0.503 (13/6)&& 0.802 & 0.117 & 0.542 (20/6)\\
\textsc{Mdiv}$^{\blacktriangle}$ & 0.720 & 0.060  & 0.460 (33/33) && 0.764 & 0.079 & 0.515 (23/16) && 0.823 & 0.096& 0.552 (29/3)\\[1pt]
\textsc{OnlyTime}$^{\diamond}$ & 0.729 & 0.068 & 0.426 (20/26) && 0.807 & 0.092 & 0.497 (26/26) && 0.826 & 0.115 & 0.534 (26/20)\\ \midrule

\textsc{HistDiv} & 0.761$^{\diamond}$ & 0.07 & 0.497$^{\blacktriangle}$ (40/13)&& 0.814 & 0.085 & 0.542$^{\blacktriangle}$ (36/26) && \textbf{0.864}$^{\ddagger}$ & 0.101 & 0.583$^{\blacktriangle}$(43/13) \\[1pt]

\textsc{HistDiv-Burst} & \textbf{0.777}$^{\diamond}$ & 0.087 & \textbf{0.509}$^{\blacktriangle}$ (33/6) && \textbf{0.830}$^{\diamond}$ & \textbf{0.113}$^{'}$ & \textbf{0.560}$^{\blacktriangle}$ (46/20) &&  0.860$^{\ddagger}$ & \textbf{0.132} & \textbf{0.601}$^{\blacktriangle}$ (43/16) \\[1pt]

\bottomrule
\end{tabular}
\caption{\textsf{SBR} at varying depths for $\Delta=$\textsf{month}. Win-Loss percentages are presented in brackets next to the $\mathcal{AT}$ scores. The superscript denotes a statistically significant difference when compared to the closest competitor (p<=0.5). For example $^{\diamond}$ represents statistically significant difference from \textsc{OnlyTime}.}

\label{tab:sbr_effect}
\end{table*}

\subsection{Setup} 

\subsubsection{Modeling Aspects and Time} 
\label{sec:setup-aspect-mining}

Since we model aspects of a documents as the entities therein we consider the named-entity tagging system~\textsc{Aida}~\cite{hoffart2011robust} for our experiments. \textsc{Aida} is the state-of-art approach for \emph{named entity disambiguation} which canonicalizes mentions of named entities (person, locations, organizations) into Wikipedia pages. Note, there are certain entities like \textsf{United States} that occur in nearly all documents and distort the performance of aspect based diversification methods. To overcome this we remove non-salient entities using an IDF filter and set the threshold to $0.2$. To extract the temporal references mentioned in the article text and populate $E(d)$ we use \emph{Tarsqi}. We consider two granularities of time intervals for the experiments -- $\Delta = \{$\textsf{month}, \textsf{year}\}.
\subsubsection{Baselines} 
\label{sub:competitors}

We evaluate the effectiveness of \textsc{HistDiv} at diversifying the search results produced by four effective classes of baselines:

\textbf{Non-temporal Baselines : } The first baseline that we considered is the standard unigram language model with dirichlet smoothing (\textsf{LM}, $\mu$ = 1000). The other approaches use the top 1000 documents returned by \textsc{LM} for diversification.  Though not designed for the historical search task we also consider \textsc{Ia-Select}~\cite{agrawal_diversifying_2009} and \textsc{PM2}~\cite{dang_diversity_2012} which are pure aspect-based diversification methods. It allows us to better highlight the nature of the task and the challenges faced when diversifying across a joint aspect-time space. 

\textbf{Temporal Diversification Baselines :} First, to create baselines more suited to the task we create temporal variants of \textsc{Ia-Select} and \textsc{Pm2} by (a) linearizing the aspects with the time window corresponding to the publication date of the document respectively denoted as \textsc{T-ia-Select} and \textsc{T-Pm2}; (b) augmenting the aspect space by including temporal aspects. Temporal aspects are represented by time-intervals and contain documents which were published in that interval. These variations are named \textsc{E-ia-Select} and \textsc{E-Pm2} respectively. Next, we consider two approaches that take time into account directly while diversifying: \textsc{OnlyTime}~\cite{lm+t+d} and \textsc{MDiv}~\cite{mdiv}. \textsc{OnlyTime}~\cite{lm+t+d} diversifies results purely in the time dimension, denoted by $\mathcal{T}_q$. \textsc{MDiv}~\cite{mdiv} is a multi-dimensional diversification approach that treats both dimensions equivalently. Finally, we also have a variant of \textsc{OnlyTime} called \textsc{EqT} which, unlike its counterpart, assumes equal distribution for the prior and does 0/1 discounting of the time windows.  We do not consider \textsc{xQuad} \cite{santos2010exploiting} due to (a) the absence of a reasonable query log for this time period and (b) the lack of clarity regarding its' adaptation without an external log for the historical search task. 

\textbf{HistDiv :} We compare both the original \textsc{HistDiv} and its burst-aware version \textsc{HistDiv-Burst} to the aforementioned baselines. For \textsc{HistDiv-Burst} we fix the the moving window size used to detect bursts to 24 months in all experiments.

To get the best performance and avoid over fitting we tune each variant for \textsf{SBR} in $\mathcal{AT}_q$ and present results using 5-fold cross validation. We evaluated all baselines for metrics mentioned in Section~\ref{sec:test-collection} at variable retrieval depths. Similar to the TREC diversity track we assumed equal distribution of subtopics for all topics.  For the temporal space $\mathcal{T}_q$ each time window in the ground truth was divided into partitions of size $\Delta$ and then given equal importance akin to \cite{lm+t+d}. Consequently, for the $\mathcal{AT}_q$ space each subtopic and relevant time interval pair is given equal importance. Note, in the joint space we do not divide the qualifying intervals into partitions. We assumed equal distribution of the qualifying intervals associated with a single subtopic.

\begin{table*}[t!]
  \footnotesize
  \centering
   \scalebox{0.95}{
  \begin{tabular}{@{}llcccccccccccccccccc@{}}\toprule

    \multicolumn{2}{l}{} &\multicolumn{2}{c}{\textsc{Ia\-P}} &&\multicolumn{2}{c}{\textsc{SBR}} &&\multicolumn{2}{c}{\textsc{NDCG}} &&\multicolumn{2}{c}{\textsc{IA-ERR}} &&\multicolumn{2}{c}{\textsc{MAP}}\\
    \midrule

    \multicolumn{2}{l}{} & M & Y && M & Y  && M & Y  && M & Y && M & Y \\ \midrule

    \multicolumn{2}{l}{\textsc{Lm}} & 0.099 & 0.099 && 0.428 & 0.428 && 0.402 & 0.402 && 0.201 & 0.201 && 0.228 & 0.228          \\

    
    \multicolumn{2}{l}{\textsc{Ia-Select}$^{\circ}$} & 0.101 & 0.101 && 0.442 & 0.442 && 0.415 & 0.415 && 0.180 & 0.180 &&0.215& 0.215 \\

    \multicolumn{2}{l}{\textsc{Pm2}$^{\star}$} &0.100& 0.100 &&0.429& 0.429 &&0.388& 0.388 &&0.213 & 0.213 &&0.241& 0.241\\

    \midrule

    \multicolumn{2}{l}{\textsc{Tia-Select}$^{\bullet}$} &\textbf{0.120}$^{\blacktriangle}$& \textbf{0.113}$^{\ddagger}$ &&0.380 & 0.361 &&\textbf{0.497}$^{\ddagger}$& \textbf{0.468}$^{\circ}$ &&0.195& 0.179 &&0.242& 0.232\\

    \multicolumn{2}{l}{\textsc{T-Pm2}$^{'}$} &0.064& 0.091 &&0.308& 0.410 &&0.232& 0.368 &&0.123& 0.176 &&0.152& 0.167\\

    \multicolumn{2}{l}{\textsc{E-Ia-Select}$^{\ddagger}$} &0.106& 0.102 &&0.435& 0.430 &&0.478& 0.412 &&0.183& 0.177 &&0.219& 0.214\\
    \multicolumn{2}{l}{\textsc{E-Pm2}$^{\dagger}$} &0.103& 0.099 &&0.422& 0.417 &&0.419& 0.379 &&0.217& 0.204 &&0.227& 0.239\\

    \multicolumn{2}{l}{\textsc{EqT}} &0.096& 0.078 && 0.441& 0.426 && 0.360& 0.331 &&0.203& 0.200 && 0.229& 0.213     \\
    
    \multicolumn{2}{l}{\textsc{Mdiv}$^{\blacktriangle}$} &0.109& 0.096 &&0.460& 0.428 &&0.389& 0.370 &&0.204& 0.203 &&0.236& 0.236 \\

    \multicolumn{2}{l}{\textsc{OnlyTime}$^{\diamond}$} &0.089& 0.076 &&0.426& 0.415 &&0.354& 0.297 &&0.196& 0.189 &&0.236& 0.220\\ \midrule

    \multicolumn{2}{l}{\textsc{HistDiv}} &0.096& 0.087 && 0.497$^{\blacktriangle}$ & 0.459$^{\circ}$&& 0.383& 0.339 && 0.229$^{\star}$& 0.208 && \textbf{0.255}$^{\bullet}$ & 0.231\\

    \multicolumn{2}{l}{\textsc{HistDiv-Burst}} &0.096& 0.096 &&\textbf{0.509}$^{\blacktriangle}$& \textbf{0.509}$^{\circ}$ &&0.375& 0.375 &&\textbf{0.231}$^{\star}$& \textbf{0.231}$^{\star}$ &&0.244 & \textbf{0.244}\\

    \bottomrule
  \end{tabular}}
  \caption{Effect of granularity $\Delta$ (M = \textsf{month}; Y = \textsf{year}) at $k=10$. The superscript denotes a statistically significant difference when compared to the closest competitor (p<=0.5). For example $^{\diamond}$ represents statistically significant difference to \textsc{OnlyTime}.}
  \label{tab:granularity_effect}
\end{table*}

\subsection{Results} 
\label{sub:subtopic_recall}

In this section we first analyze the performance of all baselines for historical query intents using subtopic recall or \textsc{SBR} since our goal is to optimize \textsc{SBR}. Even though the measure of choice for a historical search task is \textsf{SBR} in the $\mathcal{AT}$-- space, observing the component spaces $\mathcal{A}$ and $\mathcal{T}$ provides a clearer explanation of our results. Note, since each space uses a different set of subtopics, the range of values in each varies considerably.  Table~\ref{sub:subtopic_recall} summarizes the effectiveness of all baselines at $\Delta =$ \textsf{month} with respect to \textsf{SBR} in the aspect space $\mathcal{A}$, temporal space $\mathcal{T}$ and aspect-temporal space $\mathcal{AT}$ at $k= \{10,15,20 \}$. Remember that we have a nested representation of important subtopics and times in our topics definition (cf. Section~ \ref{sec:test-collection}). A document $d$ is relevant to a subtopic in the ground truth, irrespective of its corresponding time interval, is said to be relevant in aspect space $\mathcal{A}$. Similarly $d$, if published in a time interval $I$ in the ground truth, is said to be relevant in the temporal space $\mathcal{T}$ irrespective of which subtopic it is relevant to.  

First, we look at the performance of the non-temporal baselines. Not surprisingly, the diversity-unaware \textsc{Lm} fares worse than most baselines. \textsc{Ia-select} performs better than \textsc{Lm} in the aspect space $\mathcal{A}_q$ but since it does not account for time, we find that \textsc{SBR} in the temporal space $\mathcal{T}_q$ is lower compared to the temporal baselines. Consequently it performs poorly in $\mathcal{AT}_q$. The proportionality-based approach \textsc{Pm2} also performs poorly suggesting that choosing aspects proportionally is detrimental to historical query intents.

Next, we consider the temporal variants of the standard diversification approaches. We find that the linearized variant \textsc{T-Ia-select} performs consistently worse than its non-temporal counter part in all spaces across all retrieval depths. This shows that linearizing aspects with time leads to \emph{over-specification of aspects} especially for smaller time granularities. \textsc{E-Ia-select} addresses this problem leading to significantly better results in all spaces when compared to \textsc{T-Ia-select}. It also performs significantly better in $\mathcal{T}_q$ when compared to \textsc{Ia-select}. As a consequence of the inherent temporal nature of aspects alluded to earlier, the subtopic recall of \textsc{E-Ia-select} in the $\mathcal{AT}_q$ improves and is either better or comparable to \textsc{Ia-select} along with  performance in the other metrics showing marked improvement (shown in Table\ref{tab:granularity_effect}). Now we consider the other temporal baselines \textsc{OnlyTime} and \textsc{MDiv} in comparison to the temporal baselines discussed above. We observe that \textsc{MDiv} has comparable to the best performance (although not the best) in both $\mathcal{T}_q$ and $\mathcal{A}_q$ and therefore is easily the best performing temporal baseline in $\mathcal{AT}_q$. Surprisingly, for $\Delta = \textsf{month}$, \textsc{OnlyTime} does not perform as well as we expected in $\mathcal{T}_q$ even though it optimizes for temporal coverage. However, for $\Delta = $ \textsf{year}, it significantly outperforms all competitors in $\mathcal{T}_q$ including \textsc{HistDiv} (results omitted from Table~\ref{tab:granularity_effect} for space reasons) which leads us to believe that \textsc{OnlyTime} is sensitive to the underlying granularity $\Delta$. Expectedly, both \textsc{OnlyTime} and \textsc{EqT} perform worse than \textsc{MDiv}. Since subtopics need not be from mutually exclusive time intervals, \textsc{OnlyTime} struggles to cover the relevant $\mathcal{AT}_q$ space properly. \textsc{MDiv}, on the other hand, by virtue of modeling aspects (albeit independently) reconciles both dimensions better and is our closest competitor.

Lastly, we turn to \textsc{HistDiv} and its burst-aware version \textsc{HistDiv-Burst}. \textsc{HistDiv} outperforms all classes of competitors in the $\mathcal{AT}_q$ space and is significantly different from its closest competitor \textsc{MDiv} by an achieved significance level˜\cite{asl} of $p \leq 0.05$. The results vindicate our choice of mining aspects temporally and considering time period granularities as aspect proportions for $\mathcal{AT}_q$.  A key point to note is that methods which perform comparably to \textsc{HistDiv} in spaces $\mathcal{A}_q$ and $\mathcal{T}_q$ like \textsc{Ia-select} and \textsc{T-Pm2} are significantly outperformed in the space $\mathcal{AT}_q$ meaning that we achieve the right trade-off between these two seemingly conflicting yet interdependent dimensions. \textsc{HistDiv-Burst} is the best performing variant in all spaces for all depths. The improvement over \textsc{HistDiv} can be attributed to the accurate identification of important time intervals pertaining to long running events. \textsc{HistDiv-Burst} which also has the advantage of being granularity free and as evidenced by our results, is a better retrieval model to find relevant documents from important aspects during important time intervals. In Table~\ref{tab:sbr_effect} we also show the win-loss percentage of queries in $\mathcal{AT}$ compared to the baseline \textsc{LM}. We see that \textsc{HistDiv} improves a sizable portion of the workload (around 43\% at $k=20$) but more importantly under performs only in a small set of queries (16\%).

Unlike \textsc{HistDiv-Burst} the other retrieval models that incorporate time are dependent on the granularity of the time window used to discretize the time dimension. To examine the effect of granularity on the baselines we report scores in $\mathcal{AT}_q$ at $k=10$ in Table~\ref{tab:granularity_effect} using all intent-aware metrics mentioned earlier in Section~\ref{sec:test-collection} for $\Delta = \{Year, Month\}$. We find that a yearly granularity works best for certain retrieval models because, for the given workload, a smaller granularity causes temporally-aware diversification methods to \emph{over represent} relevant time periods. \textsc{HistDiv-Burst} on the other hand is more robust against over representing long running subtopics like \textsf{Giuliani's mayoralty}. However for queries like \texttt{reunification of germany}, temporal spread is not key to diversity but instead it is imperative to diversify within a small set of time intervals. Here \textsc{T-Pm2}, irrespective of granularity, is able to focus on to the dominant time window ``1989'' and covers as many aspects as possible within this window. On the other hand, \textsc{HistDiv} covers enough important aspects in the important time intervals and subsequently diversifies to find aspects in other less relevant intervals. Although with sufficient tuning \textsc{HistDiv} can perform as good or if not better than \textsc{T-Pm2}.


Interestingly, \textsc{HistDiv} also performs the best in $\mathcal{A}_q$ which leads us to believe that temporal aspects guide us to make the right choice of subtopics for temporally ambiguous queries. However this is achieved at the cost of precision. In Table~\ref{tab:granularity_effect}, we can also see the performance of all retrieval models for the other metrics. Historical search is cast as a recall-oriented task and hence it is not surprising to find \textsc{HistDiv} is not the best in \textsf{IA-P}. The baseline \textsc{Tia-Select} consistently outperforms other competitors but does so by producing aspect-redundant documents indicated by the low \textsf{SBR} score. We also notice that \textsc{Ia-Select} and \textsc{Mdiv} achieve good precision with satisfactory subtopic recall. Balancing precision and recall is key to satisfying the target user group. In the next section we choose to focus on the intended users of our search system. We first briefly highlight the quantitative results from user-centric metrics. We then take a more qualitative approach to study if this drop in precision for \textsc{HistDiv} is detrimental to user satisfaction in historical search tasks. Finally for completeness, though not included in Table~\ref{tab:sbr_effect}, we considered two other competitors -- the search engine on the official New York Times portal and a commercial search engine using the appropriate site (nytimes.com) and date filters (1987-2007) on June 9, 2015. The \textsf{SBR} scores achieved by them were the lowest -- 0.288 and 0.312 at $k=10$ respectively. Such low recall can be attributed to the fact that commercial search engines favor recency and more popular information needs reaffirming the need for specialized retrieval models designed for historical intents.

\subsection{User Study}
\label{sec:user-centric-study}
We now turn our attention to the target users to get a deeper understanding of what they perceive is a good historical overview from search results. In this section we attempt to answer the following research question: \emph{Despite a loss in precision, are users satisfied with the quality of the overview derived from \textsc{HistDiv} when compared to its competitors?}
In traditional IR, to quantify if search results are satisfactory for users (who inherently examine a result list from top to bottom) metrics such as \textsf{IA-ERR} and $\alpha$-\textsf{NDCG} are used. From Table~\ref{tab:granularity_effect} we see that \textsc{HistDiv} performs well in \textsf{IA-ERR}, which is one of the primary metrics for TREC's diversity web track, due to its ability to not only rank diverse documents higher but also cover more subtopics in the top 10 results. The lack of precision is reflected in the low \textsf{NDCG}($\alpha$ set to 0.5) scores although \textsc{HistDiv} has the highest \textsf{MAP} score. What these results do not indicate however is the overall satisfaction of the user when trying to discern an overview from the top-K search results as a whole. We see quantitatively that we cover more subtopics at important time periods but how good is the overview from a user perspective? Is the lack of precision detrimental to the overall user experience?

\textbf{Study design: } We attempt to answer the research question by comparing \textsc{HistDiv} against \textsc{Ia-Select} because \textsc{Ia-Select} exhibits high precision coupled with good recall. It is also distinctly different from \textsc{HistDiv} since it is a pure aspect based diversification algorithm. Due to the sheer number of baselines, comparing against all is prohibitive. We used a within-subjects study design to determine which approach produces a better overview. We had a total of 10 participants comprised of 3 senior historians, 1 humanities researcher and 6 computer science graduate students. The historians and the humanities researcher represent the expert opinion. Participants were required to compare and contrast both approaches for a subset of the query workload from the test collection. Of the three historians two were from the United States and the other from the United Kingdom. All of them cited previous experience using news archives for their research. The non-expert users were from three different nationalities and possessed strong proficiency in the English language. All participated voluntarily in the study.

We selected 15 topics at random from our workload and generated the top 10 results for each from both approaches. Each participant was given between 2-5 topics to evaluate. To take into account the varying familiarity of participants with their designated topics we instructed them to first read the history section of the relevant Wikipedia page for the topic. We suggested a period of 5-15 minutes for this preparation. Following this period, participants were shown the top 10 results from both methods and asked the following question : "Which result ranking gives you a better overview of the history for the given topic and why?". The approaches were anonymized and presented to the users as "X" (\textsc{Ia-Select}) and "Y" (\textsc{HistDiv}). Participants were instructed to argue their choice in the form of free text. We opted against structured questions since a good overview can be subjective and many unforeseen nuances cannot be captured. Each topic from the randomly selected subset was evaluated by 3 different participants to account for inter-rater agreement. Participants were instructed to complete each topic sequentially in no particular order. Note that the same interface was used to display both result lists. 

\textbf{Outcomes: } We obtained complete results (3 distinct raters for each topic) for 14 out of the 15 topics. The average duration was 25 minutes per topic. Participants emailed their responses individually at the end of the study. They resorted to using either bullet points or paragraphs when explaining their choices. For the evaluation, we consider an approach is superior to the other when the majority of raters vote in its favor. For the sake of readability we de-anonymize the approaches in the participants' responses in the proceeding discussion. Overall we found that participants preferred \textsc{HistDiv}'s ranking in order to get a historical overview for 10 out of 14 topics. 5 out of the 14 topics had 100\% agreement with 4 of those being in favor of \textsc{HistDiv}. The positive comments from the participants often mentioned good coverage and a better ranking of subtopics within the top 10. Some participants explicitly stated  "\textit{good diversity in results}" while others justified their choices with examples of subtopics covered exclusively by the winning approach. \textsc{HistDiv}'s span of coverage was immediately apparent to the participants. One of the topics where \textsc{HistDiv} out performed \textsc{Ia-Select} is \texttt{Bob Dole}. Participants felt that the increased coverage provided a more rounded picture when compared to \textsc{Ia-Select} ( "\textit{Histdiv mentions an article which talks about the role his wife plays which provides a more complete representation of Bob Dole}" ) showing that emphasis on recall is justified for this task. \texttt{Al Gore} was another such topic -  "\textit{HistDiv has articles about the Oscar winning movie but the other doesn't}" - indicating that an important subtopic was uncovered. The same participant also noted that "\textit{Ia-Select has 7 articles from the year 2000 -> not diverse}". This redundancy is due to over-specification of the time period caused by \textsc{Ia-Select}'s lack of temporal awareness. 

Even though participants were instructed to judge the quality of overview from the top 10 as a whole there was a tendency to be critical of the ranking. While two participants indicated that chronological ordering in the top 10 would be easier to understand, the others preferred seeing documents from the most important subtopic towards the top. For \texttt{Bob Dole}, a participant stated that "\textit{HistDiv is better as it returns his presidential campaign as the top ranked document which highlights the pinnacle of his political career}". For the topic \texttt{Oklahoma City}, an expert remarked that the results from \textsc{HistDiv} were more "\textit{on point}". He argued that while both approaches rightly picked articles about the Oklahoma city bombing at the top, \textsc{HistDiv} returned the more relevant result from the most relevant time period - "BOMB SUSPECT IS HELD, ANOTHER IDENTIFIED; TOLL HITS 65 AS HOPE FOR SURVIVORS FADES" vs "Oklahoma City, a Year Later". The former headline from \textsc{HistDiv} is a vital primary source whereas the latter from \textsc{Ia-Select} is a less relevant secondary source. Here \textsc{HistDiv}'s superior modeling of document utility helps it pick highly relevant documents from the more important historical subtopics first. This difference in importance between subtopics is not captured in the quantitative results due to our classical assumption of equally relevant subtopics. For the topics where participants agreed that \textsc{Ia-Select} did better than \textsc{HistDiv}, lack of precision was cited as the main factor. Seven participants cited the number of relevant articles as the deciding factor which is a valid concern. We found from the responses that there were two types of irrelevant articles: articles completely unrelated to the topic and articles seemingly less relevant than the others that explicitly covered important historical facts. We observed this bias towards precision almost exclusively from the non-experts although for cases like \texttt{Landon Donovan} (the former captain of the U.S. mens soccer team) both sets of users agreed that articles about politicians with similar names hurt the overview. Similarly for \texttt{Charlie Sheen} participants responded with comments such as "\textit{HistDiv includes a few articles which do not seem to be about Charlie Sheen at all or concern him only very marginally (articles ranked 2,3,7) , hard to get overview if a third of articles are not really about about Charlie Sheen}" upon encountering articles about movie listings or his father Martin Sheen. 

An interesting topic that divided opinion was \texttt{Rudolph Giuliani}. In \textsc{HistDiv}'s ranking there was an article regarding an interview on Giuliani's personal life that non-experts considered irrelevant.  An expert on the other hand acknowledged the presence of less relevant results but mentioned that for certain topics these documents are not as irrelevant as they first seem since you can find valuable contextual information -- "\textit{(For Giuliani) HistDiv is better than Ia-Select because it provides a better mix of political and personal information}". This tendency to evaluate the overview from multiple perspectives was also observed in other topics. For the topic \texttt{Atlantic City}, a participant stated that "\textit{HistDiv gives a more well argued discussion about the political choices that have been taken in order to reinvent the city}". This shows that the lack of precision is detrimental to user satisfaction although it is highly dependent on the topic. When the inherent aspect diversity of the topic is low, \textsc{HistDiv}'s tendency to increase recall ranks irrelevant documents higher. An interesting direction for future work is to design methods that allow automatic adjusting of parameters to increase precision rather than recall for certain types of queries.

From the observations in both experiments we can conclude that (a) \textsc{HistDiv} is consistently better at finding primary sources by best diversifying the aspect-time space indicated by high subtopic recall (b) \textsc{HistDiv} shows promise for pure aspect-based diversification of temporally ambiguous queries (c) In isolated cases users feel that the lack of precision hurts \textsc{HistDiv} but in the majority of cases the emphasis on recall provides a more holistic overview.

\section{Conclusion \& Outlook}

In this paper we introduced the notion of a historical query intent over longitudinal news collections like news archives. We cast the problem as diversification task in a new aspect-time space. To evaluate the task we built a new temporal test collection based on 20 years of the \emph{New York Times} collection. We introduced \textsc{HistDiv} which shows improvements over temporal and non-temporal methods for most of the time-aware diversification methods. We also outperform all competitors in subtopic recall over the joint space showing the suitability of our approach for historical query intents. We observe that \textsc{HistDiv} works well for topics which have aspects that span across multiple time intervals and have fluctuating importance at different times. It trades-off nicely between important aspects and important times which we perceive as important in historical search. \textsc{HistDiv} does not perform quite as well for queries which only one dominant aspect at a certain time window. \textsc{HistDiv} also acheives lower precision than some of its competitors although the user study showed that only in isolated cases there is a loss in overview quality. In the future, for more practical settings where training data is little to none, we want to investigate the usage of query related features like the degree of temporal variance of aspects, the number of bursts, etc. to estimate the parameters used. This opens up exciting future work opportunities to automatically identify queries of different historical intents and evaluate them accordingly.

\section{Acknowledgment}

This work was carried out under the context of the ERC Grant (339233) ALEXANDRIA. We thank Prof. Jane Winters and her colleagues from the Institute of Historical Research at the University College of London for their help and cooperation.

{ \footnotesize
\bibliographystyle{abbrv} 

\begin{thebibliography}{10}

\bibitem{ukarchive}
British newspaper archive http://www.britishnewspaperarchive.co.uk/.

\bibitem{caliarchive}
California digital newspaper collection, http://cdnc.ucr.edu/cgi-bin/cdnc.

\bibitem{nytarchive}
New york times archives, http://timesmachine.nytimes.com/browser.

\bibitem{wikiminer}
Wikiminer, http://wikipedia-miner.cms.waikato.ac.nz.

\bibitem{agrawal_diversifying_2009}
R.~Agrawal, S.~Gollapudi, A.~Halverson, and S.~Ieong.
\newblock Diversifying search results.
\newblock In {\em Proceedings of the Second {ACM} International Conference on
  Web Search and Data Mining}, {WSDM} '09, pages 5--14, New York, {NY}, {USA},
  2009. {ACM}.

\bibitem{Alonso:2007}
O.~Alonso, M.~Gertz, and R.~Baeza-Yates.
\newblock On the value of temporal information in information retrieval.
\newblock {\em SIGIR Forum}, 41(2):35--41, Dec. 2007.

\bibitem{Alonso:2009}
O.~Alonso, M.~Gertz, and R.~Baeza-Yates.
\newblock Clustering and exploring search results using timeline constructions.
\newblock In {\em Proceedings of the 18th ACM Conference on Information and
  Knowledge Management}, CIKM '09, pages 97--106, New York, NY, USA, 2009. ACM.

\bibitem{anand_index_2012}
A.~Anand, S.~Bedathur, K.~Berberich, and R.~Schenkel.
\newblock Index maintenance for time-travel text search.
\newblock In {\em Proceedings of the 35th International {ACM} {SIGIR}
  Conference on Research and Development in Information Retrieval}, {SIGIR}
  '12, pages 235--244, New York, {NY}, {USA}, 2012. {ACM}.

\bibitem{lm+t+d}
K.~Berberich and S.~Bedathur.
\newblock Temporal diversification of search results.
\newblock In {\em SIGIR 2013 Workshop on Time-aware Information Access (TAIA
  2013)}, 2013.

\bibitem{berberich_language_2010}
K.~Berberich, S.~Bedathur, O.~Alonso, and G.~Weikum.
\newblock A language modeling approach for temporal information needs.
\newblock In {\em Proceedings of the 32Nd European Conference on Advances in
  Information Retrieval}, {ECIR}'2010, pages 13--25, Berlin, Heidelberg, 2010.
  Springer-Verlag.

\bibitem{berberich_time_2007}
K.~Berberich, S.~Bedathur, T.~Neumann, and G.~Weikum.
\newblock A time machine for text search.
\newblock In {\em Proceedings of the 30th Annual International {ACM} {SIGIR}
  Conference on Research and Development in Information Retrieval}, {SIGIR}
  '07, pages 519--526, New York, {NY}, {USA}, 2007. {ACM}.

\bibitem{bingham2010digitization}
A.~Bingham.
\newblock The digitization of newspaper archives: Opportunities and challenges
  for historians.
\newblock {\em Twentieth Century British History}, 21(2):225--231, 2010.

\bibitem{brucato_metric_2014}
M.~Brucato and D.~Montesi.
\newblock Metric spaces for temporal information retrieval.
\newblock In M.~d. Rijke, T.~Kenter, A.~P.~d. Vries, C.~Zhai, F.~d. Jong,
  K.~Radinsky, and K.~Hofmann, editors, {\em Advances in Information
  Retrieval}, number 8416 in Lecture Notes in Computer Science, pages 385--397.
  Springer International Publishing, Jan. 2014.

\bibitem{campos_survey_2014}
R.~Campos, G.~Dias, A.~M. Jorge, and A.~Jatowt.
\newblock Survey of temporal information retrieval and related applications.
\newblock {\em {ACM} Comput. Surv.}, 47(2):15:1--15:41, Aug. 2014.

\bibitem{Campos:2012}
R.~Campos, A.~M. Jorge, G.~Dias, and C.~Nunes.
\newblock Disambiguating implicit temporal queries by clustering top relevant
  dates in web snippets.
\newblock In {\em Proceedings of the The 2012 IEEE/WIC/ACM International Joint
  Conferences on Web Intelligence and Intelligent Agent Technology - Volume
  01}, WI-IAT '12, pages 1--8, Washington, DC, USA, 2012. IEEE Computer
  Society.

\bibitem{carbonell1998use}
J.~Carbonell and J.~Goldstein.
\newblock The use of mmr, diversity-based reranking for reordering documents
  and producing summaries.
\newblock In {\em Proceedings of the 21st annual international ACM SIGIR
  conference on Research and development in information retrieval}, pages
  335--336. ACM, 1998.

\bibitem{Carterette:2009:PMR:1645953.1646116}
B.~Carterette and P.~Chandar.
\newblock Probabilistic models of ranking novel documents for faceted topic
  retrieval.
\newblock In {\em Proceedings of the 18th ACM Conference on Information and
  Knowledge Management}, CIKM '09, pages 1287--1296, New York, NY, USA, 2009.
  ACM.

\bibitem{case1991collection}
D.~O. Case.
\newblock The collection and use of information by some american historians: a
  study of motives and methods.
\newblock {\em The Library Quarterly}, pages 61--82, 1991.

\bibitem{choi2012temporal}
J.~Choi and W.~B. Croft.
\newblock Temporal models for microblogs.
\newblock In {\em Proceedings of the 21st ACM international conference on
  Information and knowledge management}, pages 2491--2494. ACM, 2012.

\bibitem{clarke2011nist}
C.~Clarke, N.~Craswell, I.~Soboroff, and E.~Voorhees.
\newblock Nist, overview of the trec2011 web track.
\newblock In {\em Proceedings of TREC}, pages 500--295, 2011.

\bibitem{clarke2009overview}
C.~L. Clarke, N.~Craswell, and I.~Soboroff.
\newblock Overview of the trec 2009 web track.
\newblock Technical report, DTIC Document, 2009.

\bibitem{clarke2008novelty}
C.~L. Clarke, M.~Kolla, G.~V. Cormack, O.~Vechtomova, A.~Ashkan,
  S.~B{\"u}ttcher, and I.~MacKinnon.
\newblock Novelty and diversity in information retrieval evaluation.
\newblock In {\em Proceedings of the 31st annual international ACM SIGIR
  conference on Research and development in information retrieval}, pages
  659--666. ACM, 2008.

\bibitem{dang_term_2013}
V.~Dang and B.~W. Croft.
\newblock Term level search result diversification.
\newblock In {\em Proceedings of the 36th International {ACM} {SIGIR}
  Conference on Research and Development in Information Retrieval}, {SIGIR}
  '13, pages 603--612, New York, {NY}, {USA}, 2013. {ACM}.

\bibitem{dang_diversity_2012}
V.~Dang and W.~B. Croft.
\newblock Diversity by proportionality: An election-based approach to search
  result diversification.
\newblock In {\em Proceedings of the 35th International {ACM} {SIGIR}
  Conference on Research and Development in Information Retrieval}, {SIGIR}
  '12, pages 65--74, New York, {NY}, {USA}, 2012. {ACM}.

\bibitem{Dong:2010a}
A.~Dong, Y.~Chang, Z.~Zheng, G.~Mishne, J.~Bai, R.~Zhang, K.~Buchner, C.~Liao,
  and F.~Diaz.
\newblock Towards recency ranking in web search.
\newblock In {\em Proceedings of the Third ACM International Conference on Web
  Search and Data Mining}, WSDM '10, pages 11--20, New York, NY, USA, 2010.
  ACM.

\bibitem{dong2010towards}
A.~Dong, Y.~Chang, Z.~Zheng, G.~Mishne, J.~Bai, R.~Zhang, K.~Buchner, C.~Liao,
  and F.~Diaz.
\newblock Towards recency ranking in web search.
\newblock In {\em Proceedings of the third ACM international conference on Web
  search and data mining}, pages 11--20. ACM, 2010.

\bibitem{Dong:2010b}
A.~Dong, R.~Zhang, P.~Kolari, J.~Bai, F.~Diaz, Y.~Chang, Z.~Zheng, and H.~Zha.
\newblock Time is of the essence: Improving recency ranking using twitter data.
\newblock In {\em Proceedings of the 19th International Conference on World
  Wide Web}, WWW '10, pages 331--340, New York, NY, USA, 2010. ACM.

\bibitem{mdiv}
Z.~Dou, S.~Hu, K.~Chen, R.~Song, and J.-R. Wen.
\newblock Multi-dimensional search result diversification.
\newblock In {\em Proceedings of the Fourth {ACM} International Conference on
  Web Search and Data Mining}, {WSDM} '11, pages 475--484, New York, {NY},
  {USA}, 2011. {ACM}.

\bibitem{duff2002accidentally}
W.~M. Duff and C.~A. Johnson.
\newblock Accidentally found on purpose: information-seeking behavior of
  historians in archives.
\newblock {\em The Library Quarterly}, pages 472--496, 2002.

\bibitem{Gupta:2014}
D.~Gupta and K.~Berberich.
\newblock Identifying time intervals of interest to queries.
\newblock In {\em Proceedings of the 23rd ACM International Conference on
  Conference on Information and Knowledge Management}, CIKM '14, pages
  1835--1838, New York, NY, USA, 2014. ACM.

\bibitem{hoffart2011robust}
J.~Hoffart, M.~A. Yosef, I.~Bordino, H.~F{\"u}rstenau, M.~Pinkal, M.~Spaniol,
  B.~Taneva, S.~Thater, and G.~Weikum.
\newblock Robust disambiguation of named entities in text.
\newblock In {\em Proceedings of the Conference on Empirical Methods in Natural
  Language Processing}, pages 782--792. Association for Computational
  Linguistics, 2011.

\bibitem{joho2014ntcir}
H.~Joho, A.~Jatowt, and R.~Blanco.
\newblock Ntcir temporalia: a test collection for temporal information access
  research.
\newblock In {\em Proceedings of the companion publication of the 23rd
  international conference on World wide web companion}, pages 845--850.
  International World Wide Web Conferences Steering Committee, 2014.

\bibitem{jones_temporal_2007}
R.~Jones and F.~Diaz.
\newblock Temporal profiles of queries.
\newblock {\em {ACM} Trans. Inf. Syst.}, 25(3), July 2007.

\bibitem{li2003time}
X.~Li and W.~B. Croft.
\newblock Time-based language models.
\newblock In {\em Proceedings of the twelfth international conference on
  Information and knowledge management}, pages 469--475. ACM, 2003.

\bibitem{liang_fusion_2014}
S.~Liang, Z.~Ren, and M.~de~Rijke.
\newblock Fusion helps diversification.
\newblock In {\em Proceedings of the 37th International {ACM} {SIGIR}
  Conference on Research \&\#38; Development in Information Retrieval}, {SIGIR}
  '14, pages 303--312, New York, {NY}, {USA}, 2014. {ACM}.

\bibitem{Metzler:2009}
D.~Metzler, R.~Jones, F.~Peng, and R.~Zhang.
\newblock Improving search relevance for implicitly temporal queries.
\newblock In {\em Proceedings of the 32Nd International ACM SIGIR Conference on
  Research and Development in Information Retrieval}, SIGIR '09, pages
  700--701, New York, NY, USA, 2009. ACM.

\bibitem{ecir/NguyenK14}
T.~N. Nguyen and N.~Kanhabua.
\newblock Leveraging dynamic query subtopics for time-aware search result
  diversification.
\newblock In {\em ECIR}, pages 222--234, 2014.

\bibitem{peetz2012adaptive}
M.-H. Peetz, E.~Meij, M.~de~Rijke, and W.~Weerkamp.
\newblock Adaptive temporal query modeling.
\newblock In {\em Advances in Information Retrieval}, pages 455--458. Springer,
  2012.

\bibitem{radinsky_behavioral_2013}
K.~Radinsky, K.~M. Svore, S.~T. Dumais, M.~Shokouhi, J.~Teevan, A.~Bocharov,
  and E.~Horvitz.
\newblock Behavioral dynamics on the web: Learning, modeling, and prediction.
\newblock {\em {ACM} Trans. Inf. Syst.}, 31(3):16:1--16:37, Aug. 2013.

\bibitem{santos2010exploiting}
R.~L. Santos, C.~Macdonald, and I.~Ounis.
\newblock Exploiting query reformulations for web search result
  diversification.
\newblock In {\em Proceedings of the 19th international conference on World
  wide web}, pages 881--890. ACM, 2010.

\bibitem{Setty:2010}
V.~Setty, S.~Bedathur, K.~Berberich, and G.~Weikum.
\newblock Inzeit: Efficiently identifying insightful time points.
\newblock {\em Proc. VLDB Endow.}, 3(1-2):1605--1608, Sept. 2010.

\bibitem{smith2004historians}
C.~Smith.
\newblock Historians and information.
\newblock 2004.

\bibitem{tibbo2003primarily}
H.~R. Tibbo.
\newblock Primarily history in america: How us historians search for primary
  materials at the dawn of the digital age.
\newblock {\em American Archivist}, 66(1):9--50, 2003.

\bibitem{zhou2013impact}
K.~Zhou, S.~Whiting, J.~M. Jose, and M.~Lalmas.
\newblock The impact of temporal intent variability on diversity evaluation.
\newblock In {\em Advances in Information Retrieval}, pages 820--823. Springer,
  2013.

\bibitem{zhu_learning_2014}
Y.~Zhu, Y.~Lan, J.~Guo, X.~Cheng, and S.~Niu.
\newblock Learning for search result diversification.
\newblock In {\em Proceedings of the 37th International {ACM} {SIGIR}
  Conference on Research \&\#38; Development in Information Retrieval}, {SIGIR}
  '14, pages 293--302, New York, {NY}, {USA}, 2014. {ACM}.

\end{thebibliography}

}

\end{document}